\documentclass[prl,twocolumn,amsmath,amssymb]{revtex4}

\usepackage{graphicx}
\usepackage{subfigure}
\usepackage{dcolumn}
\usepackage{array}
\usepackage{bm}
\usepackage[hang]{footmisc}

\usepackage[usenames,dvipsnames]{color}

\begin{document}

\title{\bf{Chern insulators from heavy atoms on magnetic substrates} \\[11pt] }
\author{Kevin F. Garrity and David Vanderbilt}
\affiliation{Department of Physics and Astronomy\\
Rutgers University, Piscataway, NJ 08854}

\date{\today}
\begin{abstract}
We propose searching for Chern insulators by depositing atomic
layers of elements with large spin-orbit coupling (e.g., Bi) on
the surface of a magnetic insulator.  We argue that such systems
will typically have isolated surface bands with non-zero Chern
numbers.  If these overlap in energy, a metallic surface with
large anomalous Hall conductivity (AHC) will result; if not, a
Chern-insulator state will typically occur.  Thus, our search
strategy reduces to looking for examples having the Fermi level in
a global gap extending across the entire Brillouin zone.  We
verify this search strategy and identify several candidate systems
by using first-principles calculations to compute the Chern number
and AHC of a large number of such systems on MnTe, MnSe, and EuS
surfaces.  Our search reveals several promising Chern insulators
with gaps of up to 140\,meV.
\end{abstract}


\maketitle


The discovery of the quantized conducting edge states characteristic
of the integer quantum Hall effect (IQHE), and their explanation in
terms of a bulk topological invariant known as the Chern number or
TKNN invariant \cite{tknn}, initiated a new emphasis on topology in
the theory of electronic structure.  In recent years, this trend has
accelerated enormously, extending also to materials with time-reversal
(TR) symmetry and leading to important discoveries including
two-dimensional (2D) quantum spin Hall (QSH) systems and
three-dimensional (3D) topological insulators (TI) \cite{ti_review,
ti_review2}.

Concerning systems with broken TR symmetry, it has been known
since the work of Haldane \cite{haldane} that it is possible in
principle to have an insulating magnetic material exhibiting a
non-zero Chern number, as in the IQHE, but in the absence of any
applied magnetic field.  These Chern insulators, or quantum
anomalous Hall insulators, would display many of the same
properties as IQHE systems, including robust edge states with
quantized conductance, potentially at room temperature.
Theoretically, it seems quite plausible that spin-orbit coupling
(SOC), when combined with broken TR symmetry, could allow for a
non-zero Chern number in an insulator just as it allows for
non-zero anomalous Hall conductivity in metals \cite{anom_hall,
anom_hall2, anom_hall3,qi_model}.  While some aspects of the
Chern-insulator state have been investigated
theoretically~\cite{timo,sinisa}, there are currently no
experimentally known examples of Chern insulators, and finding one
remains a major challenge in condensed matter physics.

Motivated in part by the spectacular recent progress concerning
other kinds of topological insulators, there has been a dramatic
renewal of interest recently in the search for experimental
realizations of the Chern-insulator state.  Previous experimental and theoretical
proposals for Chern insulators have typically involved starting
with non-magnetic topological-insulator or Dirac-cone systems,
such as HgTe quantum wells \cite{magdope_hgte, buhmann}, graphene
\cite{magdope_graphene}, and Bi$_2$Se$_3$ \cite{magdope_bi2se3, magdopeti,hedgehogdopebi2se3,TlBiTe2dope},
and doping them with magnetic ions in order to break TR symmetry
in such a way as to generate a Chern-insulator state~\cite{ti_thinfilm}.  
While these
proposals are promising, there are serious challenges associated
with this strategy, including the difficulties of magnetically
doping these materials in a controlled fashion, understanding the
role of the associated disorder, aligning the spins of the
dopants, and keeping these small-gap materials insulating during
the process.  Further proposals have focused on avoiding doping by
using thin layers of stoichiometric compounds like
GdBiTe$_3$~\cite{gdbite3} and
HgCr$_2$Se$_4$~\cite{hgcr2se4}, but the band gaps remain small.

\begin{figure}
\centering 
\includegraphics[width=3.5in]{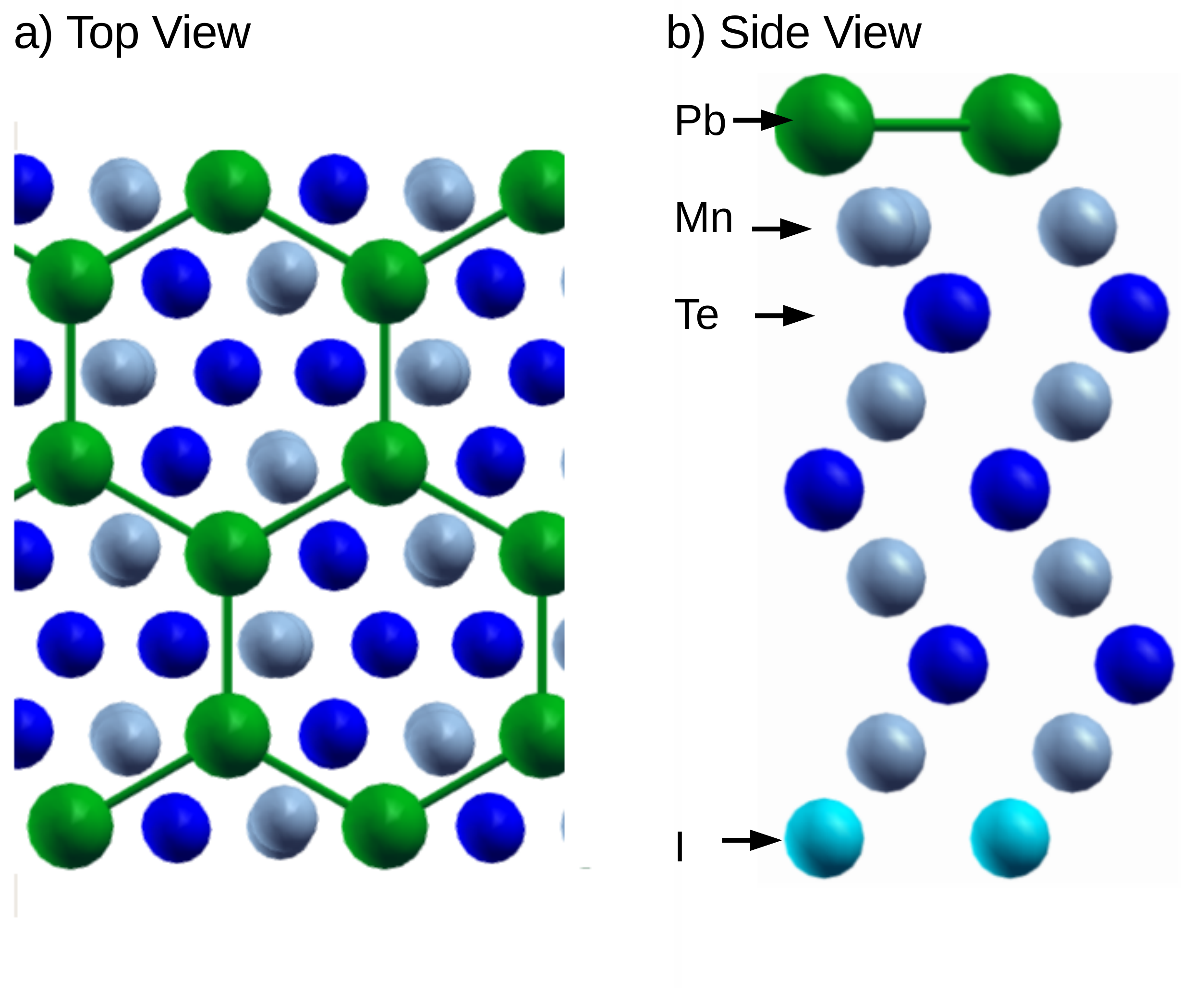}
\caption{ a) Top view and b) side view of the tripled surface
unit cell of MnTe with 2/3 ML Pb.  Pb is in green, Mn in gray,
Te is dark blue and I in cyan.}
\label{fig:cell}
\end{figure}

In this work, we propose an alternate search strategy for Chern
insulators that overcomes many of the materials challenges of
previous work, and we use first principles calculations to prove
its viability.  In addition, we suggest several candidate systems
with non-zero Chern numbers, including several with significant
gaps.  Our proposal is to start with a known magnetically-ordered
insulating substrate, choose a surface which breaks TR
symmetry, and deposit a layer of heavy atoms (Pt-Bi) with large
SOC (see Fig.~\ref{fig:cell}).  In other words,
we directly combine the two key ingredients necessary to create
a Chern insulator: broken TR symmetry and large
SOC.  By starting with a large-band-gap substrate
with naturally aligned spins, we avoid the difficulties related
to magnetic dopants and disorder.  In addition, by including the heaviest
possible atoms, we maximize the SOC and have the
potential for band gaps over 0.5\,eV.

As will be discussed shortly, our strategy
\textit{generically} gives rise to surface bands in the
bulk band gap having two crucial properties: (i) they are
\text{isolated}, in the sense that surface band $N$ does not
touch bands $N\pm 1$ anywhere in the 2D Brillouin zone
(BZ); and (ii) these bands \textit{typically carry non-zero
Chern numbers}.  Property (i) insures that the
minimum direct gap between bands $N$ and $N+1$ is generically
positive.  Property (ii) implies that if the \text{indirect}
gap between these bands is also positive, so that a global
gap exists, then a surface with $N$ filled surface
bands is likely to realize the Chern-insulator state. 
The essential challenge for our strategy, therefore, is to choose a
magnetic substrate and a heavy-adatom surface decoration that
can put the Fermi energy in a global gap of the 2D surface
bandstructure.  As we show below, this is not an insurmountable
difficulty.

We give here a brief justification of properties (i) and (ii) above,
although they are also amply illustrated by the results presented
later.  Regarding (i), it is well known that, in the absence of
special symmetries, accidental degeneracies between bands in 2D
generically do not occur.  This fact can be understood by considering
that the Hamiltonian near any potential crossing between two bands can
be written as a linear combination of the three Pauli matrices; for
the bands to cross, the coefficients of all three matrices must
vanish.  In the 2D $(k_x,k_y)$ space of the BZ, this will not happen
except by special tuning of some third parameter.  Moreover, since we
consider systems with a net magnetization at the surface in the
presence of SOC, neither time reversal (TR) nor common
crystallographic point symmetries (e.g., $3m'$ for most of the cases
below) enforce degeneracies at any high-symmetry points in the BZ.
Regarding property (ii), we find that if the SOC, magnetic exchange,
and interatomic hoppings are all of comparable magnitude, then
non-zero Chern numbers are typical.  To illustrate this, we performed
a numerical study of random tight-binding Hamiltonians consisting
of between two and six orbitals on a square lattice with random complex
hoppings to the four adjacent and four diagonal unit cells,
and found that the Chern numbers of the resulting bands appear to be normally
distributed around zero with a standard deviation of 1-2.  In other
words, non-zero Chern numbers are abundant in 2D in the absence
of TR or other special symmetries.

In order to prove the viability of our proposal, we perform
first-principles
calculations on surfaces of three insulating magnetic substrates
with a variety of monolayer or submonolayer heavy-atom coverages,
searching for promising Chern-insulator systems.
We consider the (001) surfaces of MnSe and MnTe in the NiAs
structure ($P6_3/mmc$), and the (111) surface of EuS in the
rocksalt structure ($Fm\bar{3}m$).  These hexagonal surfaces are
closely related, as the structures only differ by the stacking
sequence of the hexagonal close-packed atomic layers.  MnSe and
MnTe are A-type antiferromagnets, with layers of ferromagnetically
aligned spins lying in the $xy$ plane, while EuS
is ferromagnetic.  For MnSe, we calculate that the spins
can be aligned in the $\pm z$-direction by applying
an epitaxial strain of $-2$\%, so we adopt this strain state in what
follows.
These substrates were chosen for their large magnetizations,
large band gaps, and simple structures.

Our computational supercells consist of slabs of four layers of the
magnetic substrate, passivated by iodine atoms on the bottom surface
and stacked with a vacuum separation of $\sim$12\,\AA, as illustrated in
Fig.~\ref{fig:cell}.
The top surfaces of the substrates are terminated on the
magnetic-atom layer (Mn or Eu) so that when the heavy atoms
(Pt through Bi) are absorbed, the direct contact with the magnetic ions
will maximizing the
exchange splitting.  We note that these terminations are all polar,
with each 1$\times$1 area donating one electron to the surface
adatoms.  This fact highlights the difficulty in combining heavy
atoms directly with magnetic atoms in a thermodynamically stable
way, as both types of atom typically prefer positive oxidation
states.  However, it may be possible to stabilize these surfaces
as metastable states. Further theoretical work, combined with
experimental investigations, will be necessary to identify which
structures can be achieved in practice.

Our first-principles plane-wave calculations are carried out
in the context of density functional theory (DFT) \cite{hk,ks}
using the local-density
approximation (LDA) \cite{lda, lda1} and the PBE
generalized-gradient approximation (GGA)~\cite{pbe} for
calculations on Mn and Eu compounds respectively.  We add
Hubbard $U$ (DFT+U) \cite{ldaplusU, ldaplusU_simplified} corrections
to Mn and Eu using literature values ($U$=5 and 6\,eV respectively)
which which have been shown to be needed to describe the bulk materials
as insulators \cite{mn_calco, eu_calco}.  We have tested
the sensitivity of our results to the choice of $U$; for variations of
1-2\,eV the calculated Chern numbers are constant, but the
magnitude of the gaps can change.  The calculations are done with
two codes: Quantum Espresso \cite{QE}, using fully-relativistic
norm-conserving non-local pseudopotentials from the OPIUM
package \cite{designed-nonlocal,opium}, as well as VASP \cite{vasp,
vasp2} using PAWs  \cite{paw, paw2} (we find very similar results
with both codes).

Results from both codes are used as input to construct
maximally-localized Wannier functions (MLWF) using WANNIER90
\cite{mlwf, wannier90}.  Chern numbers and band gaps are
calculated using Wannier interpolation of the band structure; the
Chern numbers are computed by sampling a dense k-point grid (from
32$\times$32 to 128$\times$128, if necessary for convergence) and
adding up the Berry phases around the loops formed by each set of
four adjacent k-points in the Brillouin zone.  We label the Chern
numbers of each gap, defined as the total Chern number of all
bands below the gap; the Chern number of an isolated band is then
just the difference between the gap Chern numbers above and below
it.  In addition, we calculate the anomalous Hall conductivity
(AHC) for a few examples using a 128$\times$128 k-point grid with
the WANNIER90 postprocessing code
\cite{qah_wan,qah_fermi,ivo_private}.

We begin our search by considering full monolayers of heavy atoms
on MnTe surfaces.  Promisingly, the bands show both large exchange
splittings and strong SOC effects, but the bandwidths
of the surface bands are about 3$~$eV, making it difficult to
find systems with gaps.  In order to reduce the dispersion of the
surface bands to the same magnitude as the spin-orbit and exchange
splittings, we consider tripled unit cells with only 1/3 or 2/3 ML
of heavy atoms adsorbed.  These structures,
which produce flatter surface bands, are the focus of the remainder
of this work.

\begin{figure}
\centering 
\includegraphics[width=3.5in]{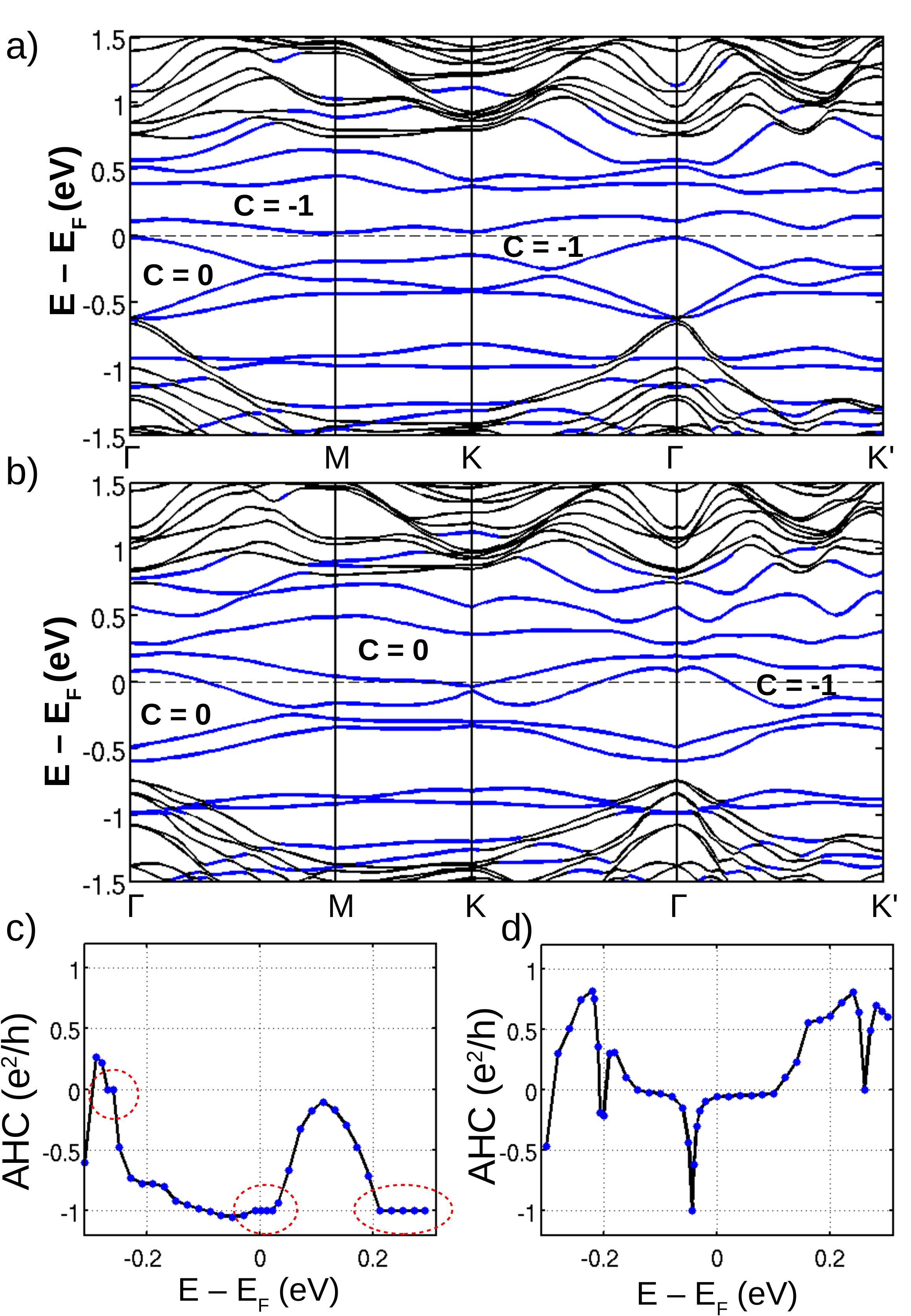}
\caption{\label{fig:pbpb} (a-b) Band structure of 2/3 Pb on MnTe
with spins in (a) the $\pm z$-direction, and (b) the $\pm x$-direction. Bands
with the most Pb character are highlighted in thick (blue) lines;
bulk-like MnTe bands are in thin (black) lines.  Chern numbers
of gaps near the Fermi level are labeled.  (c-d) Anomalous Hall
conductivity of band structures in (a-b), in units of $e^2/h$.
Quantized plateaus are highlighted.}
\end{figure}

In Figs.~\ref{fig:pbpb}(a-b) we show the band structure of 2/3 ML of
Pb of MnTe arranged in a honeycomb lattice
(see Fig.~\ref{fig:cell}) with the spins aligned
along the $\pm z$ and $\pm x$ direction respectively.  The system
with spins in the $\pm z$ direction is a Chern insulator ($C$=$-$1) with a
direct band gap of 126 meV and an indirect band gap of 36 meV,
providing a proof of principle for our search method.
The effect of the non-zero Chern numbers on the AHC can be
seen for the insulating case of Fig.~\ref{fig:pbpb}(c),
which shows quantized Hall plateaus (circled).  For the case of
a semimetal, Fig.~\ref{fig:pbpb}(d), the plateaus disappear but the
AHC remains on the order of $e^2/h$, especially near avoided crossings.
Incidentally, we find a similar behavior for the case of a
honeycomb layer of Pb atoms at the same lattice constant in
vacuum
with a Zeeman field of about 0.5\,eV representing the coupling
to Mn; such a (hypothetical) system also produces similar flat
bands with non-zero Chern numbers.

The Pb honeycomb structure has seven Pb $p$ bands occupied
in the tripled unit cell.  We can change the doping of the
system, modifying this number down to one and up to nine,
by scanning through the (Au, Hg, Tl, Pb, Bi) series of
heavy elements.  We have carried out corresponding DFT
calculations of all of these systems, and the results are
summarized in Table \ref{table:chern}.  We find that
2/3 ML Au on MnTe with spins along $z$ also produces a Chern
insulator with properties similar to that of Pb, except that
the Chern number is now $C$=+1.

\begin{table} 
\begin{center} 
\begin{ruledtabular} 
\begin{tabular}{llcrrr} 
 Substrate & Surface & Spin      & $C$ &
 \multicolumn{1}{c}{$E_{\rm g}^{\rm dir}$} & 
 \multicolumn{1}{c}{$E_{\rm g}^{\rm indir}$} \\
  &  & direction & & \multicolumn{1}{c}{(meV)} & \multicolumn{1}{c}{(meV)} \\
\hline
MnTe       & \textbf{AuAu} & $z$ & 1     & 141  & 36 \\
MnTe       &  AuAu         & $x$ & $m$   & $m$  & $m$ \\
MnTe       &  HgHg         & $z$ & 0     &   31 & $-$341 \\
MnTe       &  TlTl         & $z$ & $m$   & $m$  & $m$ \\
MnTe       &  \textbf{PbPb}& $z$ &  $-$1 &  126 &   36 \\
MnTe       &  PbPb         & $x$ &  $-$1 &  12  & $-$156 \\
MnTe       &  BiBi         & $z$ & $m$   &  $m$ & $m$ \\
\hline
MnSe       &  Pb           & $z$ &     0 &  25  & 24 \\
MnSe       &  AuAu         & $z$ &     1 &  64  & $-$731 \\
MnSe       &  \textbf{PbPb}& $z$ &  $-$1 & 213  & 1 \\
MnSe       &  PbPb         & $x$ &  $-$1 &  12  & $-$103 \\
MnSe       &  PbBi         & $z$ &  $-$2 &  31  & $-$9 \\
MnSe       &  \textbf{PbPbI}& $z$ & $-$3 &  84  &   56 \\
MnSe       &  \textbf{BiI} & $z$ &   1   &  302 &    41 \\
MnSe       &  \textbf{BiBr}& $z$ &   1   &  213 &   142 \\
MnSe       &  TlI          & $z$ &   0   &    5 &  $-$53 \\
MnSe       &  HgSe         & $z$ & $-$1  & 22 & $-$23\\
\hline
EuS       &  PbPb         & $z$ & $-$1 & 91 & $-$48 \\
EuS       &  AuAu         & $z$ & 0 & 188 & $-$251 \\
\end{tabular} 
\end{ruledtabular} 
\caption{Chern number ($C$) and direct ($E_{\rm g}^{\rm dir}$) and
indirect ($E_{\rm g}^{\rm indir}$) gaps for surface adatoms on four layers of
magnetic insulator.  Surfaces are labeled by number of adatoms
per tripled unit cell (e.g., PbPbI consists of 2/3 ML Pb and 1/3 ML I).
Chern insulators are in bold.  Systems with $E_{\rm g}^{\rm
dir}<2$\,meV are labeled $m$ (``metallic'').}
\label{table:chern} 
\end{center} 
\end{table} 

While encouraging, our initial examples of Chern insulators
on MnTe have two problems: (i) the band gaps are rather
small, and (ii) experimentally the spins of bulk MnTe lie in
the $xy$ plane, which we find tends to close the gaps.
We address these problems by searching
though a variety of structures on EuS and strained MnSe surfaces,
which have their spins in the $z$ (surface-normal) direction.
We consider tripled surface unit cells decorated with 1-2 heavy
adatoms and
0-2 electron-accepting adatoms (S, Se, I, Br), thus allowing us
to tune the filling of the surface bands towards potentially
Chern-insulating combinations.  We perform 51 calculations in our
initial search by using slabs of magnetic insulator which are only
two layers thick.
We find that 30 of these are metallic, 6 because of a near-zero
direct gap and 24 because of indirect overlap
($E_{\rm g}^{\rm indir}<0$). Of the latter, the
Chern number of the gap is non-zero for 14, of which 3 are
nearly Chern insulators ($E_{\rm g}^{\rm indir}>-10$\,meV).
Of the remaining surfaces, 11 are trivial insulators and 9
are Chern insulators (4 of which, however, have
$E_{\rm g}^{\rm indir}<10$\,meV).
Then we re-calculate several interesting candidate systems
using four-layer slabs, usually finding similar surface bands but
with modified gaps.  The results for some representative cases
are presented in Table \ref{table:chern}.  In general,
the systems with EuS and MnSe are similar, with both producing many
non-zero Chern numbers.  On the other hand, fewer of the EuS systems
are insulating; we trace this to the
reduced overlap between the very localized Eu $4f$-states
and the surface adatoms, leading to reduced exchange splittings.

As shown in Table \ref{table:chern}, we find a variety of materials
with non-trivial Chern numbers, including six Chern insulators with
indirect gaps ranging from a minuscule 1\,meV to a robust 142\,meV
(BiBr on MnSe).  To our knowledge, this is the largest calculated
band gap for a theoretically proposed Chern-insulator state.
In addition, BiI on MnSe has a direct
band gap which never falls below 302 meV, suggesting that there is
no fundamental principle which prohibits this search method from
producing gaps at least as large as those seen in Bi$_2$Se$_3$,
the largest-gap topological insulator.

\begin{figure}
\centering 
\includegraphics[width=3.5in]{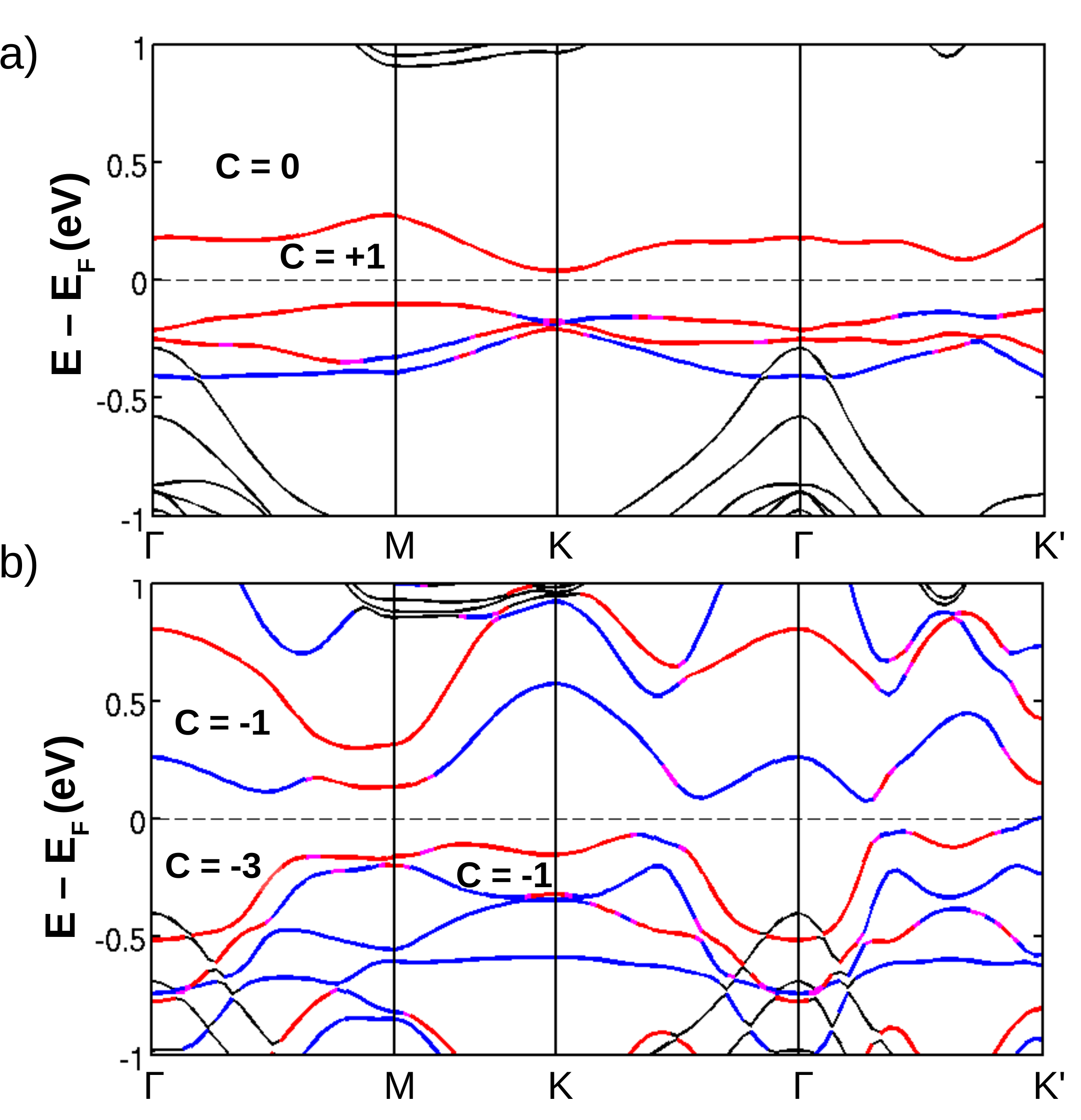}
\caption{\label{fig:good} Band structures of atoms on $-$2\%
epitaxially strained MnSe.  Top: 1/3 ML Br, 1/3 ML Bi, Bottom: 2/3 ML Pb,
1/3 ML I.  Majority spin-up surface bands are in red; majority
spin-down surface bands are in blue; bulk-like bands are in thin
black lines.  Chern numbers near the Fermi level are labeled.}
\end{figure}

Fig.~\ref{fig:good} shows
the band structures of two of our largest-gap Chern insulators,
2/3 ML Pb plus 1/3 ML I (PbPbI) and 1/3 ML Bi plus 1/3 ML Br (BiBr)
on strained MnSe.  These two band structures are typical of these
systems, except that these examples have unusually large gaps at the
Fermi level.  Systems with one heavy atom per tripled unit cell,
like BiBr, frequently have flat bands with large gaps; however,
many of these gaps are trivial.   In contrast, systems with two
heavy atoms, like PbPbI, have more non-zero Chern numbers but
fewer large gaps; this results from the larger number of dispersive bands
around the Fermi level, which leads to more avoided crossings.

While we have not addressed the kinetic or thermodynamic
stability of these particular Chern-insulator surface
systems under realistic
experimental conditions, our strategy of putting heavy atoms on
the surfaces of magnetic substrates produces so many non-trivial
band structures that it is likely that suitable candidates can be
achieved through future collaboration between theory and experiment.
In addition, it should be possible to search this class of surfaces
experimentally for candidates that are metallic but have
a large AHC, on the order of $e^2/h$ as in Fig.~\ref{fig:pbpb}(d).
Since this implies that strong SOC is already generating a
non-trivial band structure, one could then attempt to tune these
candidates via chemical
additions or substitutions, strain, or spin-alignment engineering,
in such a way as to arrive at a Chern-insulating state.
Undoubtedly, a combined experimental and theoretical search
will be the best strategy for arriving at the desired
Chern-insulating surfaces.

In summary, we have proposed a search strategy for finding Chern
insulators which avoids many of the materials-related difficulties of
previous proposals.  Our approach is to use magnetically insulating
large-gap substrates and decorate their surfaces with elements having
large spin-orbit coupling.  We have demonstrated the viability of
this approach with first-principles calculations, finding
a number of Chern insulators with gaps on the order of
50-150\,meV, and have discussed the possibilities for the
experimental realization of this strategy.

\vspace{0.3cm}
\noindent{\bf Acknowledgments}
\vspace{0.3cm}

This work was supported by NSF Grant DMR-10-05838. We thank
Ivo Souza for providing a beta version of the WANNIER90 code,
and Sinisa Coh for useful discussions.


\begin{thebibliography}{41}
\expandafter\ifx\csname natexlab\endcsname\relax\def\natexlab#1{#1}\fi
\expandafter\ifx\csname bibnamefont\endcsname\relax
  \def\bibnamefont#1{#1}\fi
\expandafter\ifx\csname bibfnamefont\endcsname\relax
  \def\bibfnamefont#1{#1}\fi
\expandafter\ifx\csname citenamefont\endcsname\relax
  \def\citenamefont#1{#1}\fi
\expandafter\ifx\csname url\endcsname\relax
  \def\url#1{\texttt{#1}}\fi
\expandafter\ifx\csname urlprefix\endcsname\relax\def\urlprefix{URL }\fi
\providecommand{\bibinfo}[2]{#2}
\providecommand{\eprint}[2][]{\url{#2}}

\bibitem[{\citenamefont{Thouless et~al.}(1982)\citenamefont{Thouless, Kohmoto,
  Nightingale, and den Nijs}}]{tknn}
\bibinfo{author}{\bibfnamefont{D.~J.} \bibnamefont{Thouless}},
  \bibinfo{author}{\bibfnamefont{M.}~\bibnamefont{Kohmoto}},
  \bibinfo{author}{\bibfnamefont{M.~P.} \bibnamefont{Nightingale}},
  \bibnamefont{and} \bibinfo{author}{\bibfnamefont{M.}~\bibnamefont{den Nijs}},
  \bibinfo{journal}{Phys. Rev. Lett.} \textbf{\bibinfo{volume}{49}},
  \bibinfo{pages}{405} (\bibinfo{year}{1982}).

\bibitem[{\citenamefont{Hasan and Kane}(2010)}]{ti_review}
\bibinfo{author}{\bibfnamefont{M.~Z.} \bibnamefont{Hasan}} \bibnamefont{and}
  \bibinfo{author}{\bibfnamefont{C.~L.} \bibnamefont{Kane}},
  \bibinfo{journal}{Rev. Mod. Phys.} \textbf{\bibinfo{volume}{82}},
  \bibinfo{pages}{3045} (\bibinfo{year}{2010}).

\bibitem[{\citenamefont{Qi and Zhang}(2011)}]{ti_review2}
\bibinfo{author}{\bibfnamefont{X.-L.} \bibnamefont{Qi}} \bibnamefont{and}
  \bibinfo{author}{\bibfnamefont{S.-C.} \bibnamefont{Zhang}},
  \bibinfo{journal}{Rev. Mod. Phys.} \textbf{\bibinfo{volume}{1057}},
  \bibinfo{pages}{1057} (\bibinfo{year}{2011}).

\bibitem[{\citenamefont{Haldane}(1988)}]{haldane}
\bibinfo{author}{\bibfnamefont{F.~D.~M.} \bibnamefont{Haldane}},
  \bibinfo{journal}{Phys. Rev. Lett.} \textbf{\bibinfo{volume}{61}},
  \bibinfo{pages}{2015} (\bibinfo{year}{1988}).

\bibitem[{\citenamefont{Jungwirth et~al.}(2002)\citenamefont{Jungwirth, Niu,
  and MacDonald}}]{anom_hall}
\bibinfo{author}{\bibfnamefont{T.}~\bibnamefont{Jungwirth}},
  \bibinfo{author}{\bibfnamefont{Q.}~\bibnamefont{Niu}}, \bibnamefont{and}
  \bibinfo{author}{\bibfnamefont{A.~H.} \bibnamefont{MacDonald}},
  \bibinfo{journal}{Phys. Rev. Lett.} \textbf{\bibinfo{volume}{88}},
  \bibinfo{pages}{207208} (\bibinfo{year}{2002}).

\bibitem[{\citenamefont{Haldane}(2004)}]{anom_hall3}
\bibinfo{author}{\bibfnamefont{F.~D.~M.} \bibnamefont{Haldane}},
  \bibinfo{journal}{Phys. Rev. Lett.} \textbf{\bibinfo{volume}{93}},
  \bibinfo{pages}{206602} (\bibinfo{year}{2004}).

\bibitem[{\citenamefont{Qi et~al.}(2006)\citenamefont{Qi, Wu, and
  Zhang}}]{qi_model}
\bibinfo{author}{\bibfnamefont{X.-L.} \bibnamefont{Qi}},
  \bibinfo{author}{\bibfnamefont{Y.-S.} \bibnamefont{Wu}}, \bibnamefont{and}
  \bibinfo{author}{\bibfnamefont{S.-C.} \bibnamefont{Zhang}},
  \bibinfo{journal}{Phys. Rev. B} \textbf{\bibinfo{volume}{74}},
  \bibinfo{pages}{085308} (\bibinfo{year}{2006}).

\bibitem[{\citenamefont{Fang et~al.}(2003)\citenamefont{Fang, Nagaosa,
  Takahashi, Asamitsu, Mathieu, Ogasawara, Yamada, Kawasaki, Tokura, and
  Terakura}}]{anom_hall2}
\bibinfo{author}{\bibfnamefont{Z.}~\bibnamefont{Fang}},
  \bibinfo{author}{\bibfnamefont{N.}~\bibnamefont{Nagaosa}},
  \bibinfo{author}{\bibfnamefont{K.~S.} \bibnamefont{Takahashi}},
  \bibinfo{author}{\bibfnamefont{A.}~\bibnamefont{Asamitsu}},
  \bibinfo{author}{\bibfnamefont{R.}~\bibnamefont{Mathieu}},
  \bibinfo{author}{\bibfnamefont{T.}~\bibnamefont{Ogasawara}},
  \bibinfo{author}{\bibfnamefont{H.}~\bibnamefont{Yamada}},
  \bibinfo{author}{\bibfnamefont{M.}~\bibnamefont{Kawasaki}},
  \bibinfo{author}{\bibfnamefont{Y.}~\bibnamefont{Tokura}}, \bibnamefont{and}
  \bibinfo{author}{\bibfnamefont{K.}~\bibnamefont{Terakura}},
  \bibinfo{journal}{Science} \textbf{\bibinfo{volume}{302}},
  \bibinfo{pages}{92} (\bibinfo{year}{2003}).

\bibitem[{\citenamefont{Thonhauser and Vanderbilt}(2006)}]{timo}
\bibinfo{author}{\bibfnamefont{T.}~\bibnamefont{Thonhauser}} \bibnamefont{and}
  \bibinfo{author}{\bibfnamefont{D.}~\bibnamefont{Vanderbilt}},
  \bibinfo{journal}{Phys. Rev. B} \textbf{\bibinfo{volume}{74}},
  \bibinfo{pages}{235111} (\bibinfo{year}{2006}).

\bibitem[{\citenamefont{Coh and Vanderbilt}(2009)}]{sinisa}
\bibinfo{author}{\bibfnamefont{S.}~\bibnamefont{Coh}} \bibnamefont{and}
  \bibinfo{author}{\bibfnamefont{D.}~\bibnamefont{Vanderbilt}},
  \bibinfo{journal}{Phys. Rev. Lett.} \textbf{\bibinfo{volume}{102}},
  \bibinfo{pages}{107603} (\bibinfo{year}{2009}).

\bibitem[{\citenamefont{Liu et~al.}(2008)\citenamefont{Liu, Qi, Dai, Fang, and
  Zhang}}]{magdope_hgte}
\bibinfo{author}{\bibfnamefont{C.-X.} \bibnamefont{Liu}},
  \bibinfo{author}{\bibfnamefont{X.-L.} \bibnamefont{Qi}},
  \bibinfo{author}{\bibfnamefont{X.}~\bibnamefont{Dai}},
  \bibinfo{author}{\bibfnamefont{Z.}~\bibnamefont{Fang}}, \bibnamefont{and}
  \bibinfo{author}{\bibfnamefont{S.-C.} \bibnamefont{Zhang}},
  \bibinfo{journal}{Phys. Rev. Lett.} \textbf{\bibinfo{volume}{101}},
  \bibinfo{pages}{146802} (\bibinfo{year}{2008}).

\bibitem[{\citenamefont{Buhmann}(2012)}]{buhmann}
\bibinfo{author}{\bibfnamefont{H.}~\bibnamefont{Buhmann}},
  \bibinfo{journal}{APS March Meeting abstract} \textbf{\bibinfo{volume}{57}},
  \bibinfo{pages}{P27.00001} (\bibinfo{year}{2012}).

\bibitem[{\citenamefont{Qiao et~al.}(2010)\citenamefont{Qiao, Yang, Feng, Tse,
  Ding, Yao, Wang, and Niu}}]{magdope_graphene}
\bibinfo{author}{\bibfnamefont{Z.}~\bibnamefont{Qiao}},
  \bibinfo{author}{\bibfnamefont{S.~A.} \bibnamefont{Yang}},
  \bibinfo{author}{\bibfnamefont{W.}~\bibnamefont{Feng}},
  \bibinfo{author}{\bibfnamefont{W.-K.} \bibnamefont{Tse}},
  \bibinfo{author}{\bibfnamefont{J.}~\bibnamefont{Ding}},
  \bibinfo{author}{\bibfnamefont{Y.}~\bibnamefont{Yao}},
  \bibinfo{author}{\bibfnamefont{J.}~\bibnamefont{Wang}}, \bibnamefont{and}
  \bibinfo{author}{\bibfnamefont{Q.}~\bibnamefont{Niu}},
  \bibinfo{journal}{Phys. Rev. B.} \textbf{\bibinfo{volume}{82}},
  \bibinfo{pages}{161414} (\bibinfo{year}{2010}).

\bibitem[{\citenamefont{Yu et~al.}(2010)\citenamefont{Yu, Zhang, Zhang, Zhang,
  Dai, and Fang}}]{magdope_bi2se3}
\bibinfo{author}{\bibfnamefont{R.}~\bibnamefont{Yu}},
  \bibinfo{author}{\bibfnamefont{W.}~\bibnamefont{Zhang}},
  \bibinfo{author}{\bibfnamefont{H.-J.} \bibnamefont{Zhang}},
  \bibinfo{author}{\bibfnamefont{S.-C.} \bibnamefont{Zhang}},
  \bibinfo{author}{\bibfnamefont{X.}~\bibnamefont{Dai}}, \bibnamefont{and}
  \bibinfo{author}{\bibfnamefont{Z.}~\bibnamefont{Fang}},
  \bibinfo{journal}{Science} \textbf{\bibinfo{volume}{329}},
  \bibinfo{pages}{61} (\bibinfo{year}{2010}).

\bibitem[{\citenamefont{Kou et~al.}(2012)\citenamefont{Kou, Jiang, Lang, Xiu,
  He, Wang, Wang, Yu, Fedorov, Zhang et~al.}}]{magdopeti}
\bibinfo{author}{\bibfnamefont{X.~F.} \bibnamefont{Kou}},
  \bibinfo{author}{\bibfnamefont{W.~J.} \bibnamefont{Jiang}},
  \bibinfo{author}{\bibfnamefont{M.~R.} \bibnamefont{Lang}},
  \bibinfo{author}{\bibfnamefont{F.~X.} \bibnamefont{Xiu}},
  \bibinfo{author}{\bibfnamefont{L.}~\bibnamefont{He}},
  \bibinfo{author}{\bibfnamefont{Y.}~\bibnamefont{Wang}},
  \bibinfo{author}{\bibfnamefont{Y.}~\bibnamefont{Wang}},
  \bibinfo{author}{\bibfnamefont{X.~X.} \bibnamefont{Yu}},
  \bibinfo{author}{\bibfnamefont{A.~V.} \bibnamefont{Fedorov}},
  \bibinfo{author}{\bibfnamefont{P.}~\bibnamefont{Zhang}},
  \bibnamefont{et~al.}, \bibinfo{journal}{Journal of Applied Physics}
  \textbf{\bibinfo{volume}{112}}, \bibinfo{pages}{063912}
  (\bibinfo{year}{2012}).

\bibitem[{\citenamefont{Xu et~al.}(2012)\citenamefont{Xu, Neupane, Liu, Zhang,
  Richardella, Wray, Alidoust, Leandersson, Balasubramanian, Sánchez-Barriga
  et~al.}}]{hedgehogdopebi2se3}
\bibinfo{author}{\bibfnamefont{S.-Y.} \bibnamefont{Xu}},
  \bibinfo{author}{\bibfnamefont{M.}~\bibnamefont{Neupane}},
  \bibinfo{author}{\bibfnamefont{C.}~\bibnamefont{Liu}},
  \bibinfo{author}{\bibfnamefont{D.}~\bibnamefont{Zhang}},
  \bibinfo{author}{\bibfnamefont{A.}~\bibnamefont{Richardella}},
  \bibinfo{author}{\bibfnamefont{L.}~\bibnamefont{Wray}},
  \bibinfo{author}{\bibfnamefont{N.}~\bibnamefont{Alidoust}},
  \bibinfo{author}{\bibfnamefont{M.}~\bibnamefont{Leandersson}},
  \bibinfo{author}{\bibfnamefont{T.}~\bibnamefont{Balasubramanian}},
  \bibinfo{author}{\bibfnamefont{J.}~\bibnamefont{Sánchez-Barriga}},
  \bibnamefont{et~al.}, \bibinfo{journal}{Nat. Phys.}
  \textbf{\bibinfo{volume}{8}} (\bibinfo{year}{2012}).

\bibitem[{\citenamefont{Niu et~al.}(2011)\citenamefont{Niu, Dai, Yu, Guo, Ma,
  and Huang}}]{TlBiTe2dope}
\bibinfo{author}{\bibfnamefont{C.}~\bibnamefont{Niu}},
  \bibinfo{author}{\bibfnamefont{Y.}~\bibnamefont{Dai}},
  \bibinfo{author}{\bibfnamefont{L.}~\bibnamefont{Yu}},
  \bibinfo{author}{\bibfnamefont{M.}~\bibnamefont{Guo}},
  \bibinfo{author}{\bibfnamefont{Y.}~\bibnamefont{Ma}}, \bibnamefont{and}
  \bibinfo{author}{\bibfnamefont{B.}~\bibnamefont{Huang}},
  \bibinfo{journal}{Applied Physics Letters} \textbf{\bibinfo{volume}{99}},
  \bibinfo{pages}{142502} (\bibinfo{year}{2011}).

\bibitem[{\citenamefont{Jiang et~al.}(2012)\citenamefont{Jiang, Qiao, Liu, and
  Niu}}]{ti_thinfilm}
\bibinfo{author}{\bibfnamefont{H.}~\bibnamefont{Jiang}},
  \bibinfo{author}{\bibfnamefont{Z.}~\bibnamefont{Qiao}},
  \bibinfo{author}{\bibfnamefont{H.}~\bibnamefont{Liu}}, \bibnamefont{and}
  \bibinfo{author}{\bibfnamefont{Q.}~\bibnamefont{Niu}},
  \bibinfo{journal}{Phys. Rev. B} \textbf{\bibinfo{volume}{85}},
  \bibinfo{pages}{045445} (\bibinfo{year}{2012}).

\bibitem[{\citenamefont{Zhang et~al.}(2012)\citenamefont{Zhang, Zhang, and
  Zhang}}]{gdbite3}
\bibinfo{author}{\bibfnamefont{H.-J.} \bibnamefont{Zhang}},
  \bibinfo{author}{\bibfnamefont{X.}~\bibnamefont{Zhang}}, \bibnamefont{and}
  \bibinfo{author}{\bibfnamefont{S.-C.} \bibnamefont{Zhang}},
  \bibinfo{journal}{arXiv:1108.4857}  (\bibinfo{year}{2012}).

\bibitem[{\citenamefont{Xu et~al.}(2011)\citenamefont{Xu, Weng, Wang, Dai, and
  Fang}}]{hgcr2se4}
\bibinfo{author}{\bibfnamefont{G.}~\bibnamefont{Xu}},
  \bibinfo{author}{\bibfnamefont{H.}~\bibnamefont{Weng}},
  \bibinfo{author}{\bibfnamefont{Z.}~\bibnamefont{Wang}},
  \bibinfo{author}{\bibfnamefont{X.}~\bibnamefont{Dai}}, \bibnamefont{and}
  \bibinfo{author}{\bibfnamefont{Z.}~\bibnamefont{Fang}},
  \bibinfo{journal}{Phys. Rev. Lett.} \textbf{\bibinfo{volume}{107}},
  \bibinfo{pages}{186806} (\bibinfo{year}{2011}).

\bibitem[{\citenamefont{Hohenberg and Kohn}(1964)}]{hk}
\bibinfo{author}{\bibfnamefont{P.}~\bibnamefont{Hohenberg}} \bibnamefont{and}
  \bibinfo{author}{\bibfnamefont{W.}~\bibnamefont{Kohn}},
  \bibinfo{journal}{Phys.\ Rev.} \textbf{\bibinfo{volume}{136}},
  \bibinfo{pages}{B864} (\bibinfo{year}{1964}).

\bibitem[{\citenamefont{Kohn and Sham}(1965)}]{ks}
\bibinfo{author}{\bibfnamefont{W.}~\bibnamefont{Kohn}} \bibnamefont{and}
  \bibinfo{author}{\bibfnamefont{L.}~\bibnamefont{Sham}},
  \bibinfo{journal}{Phys.\ Rev.} \textbf{\bibinfo{volume}{140}},
  \bibinfo{pages}{A1133} (\bibinfo{year}{1965}).

\bibitem[{\citenamefont{Ceperley and Alder}(1980)}]{lda}
\bibinfo{author}{\bibfnamefont{D.~M.} \bibnamefont{Ceperley}} \bibnamefont{and}
  \bibinfo{author}{\bibfnamefont{B.~J.} \bibnamefont{Alder}},
  \bibinfo{journal}{Phys. Rev. Lett.} \textbf{\bibinfo{volume}{45}},
  \bibinfo{pages}{566} (\bibinfo{year}{1980}).

\bibitem[{\citenamefont{Perdew and Zunger}(1981)}]{lda1}
\bibinfo{author}{\bibfnamefont{J.~P.} \bibnamefont{Perdew}} \bibnamefont{and}
  \bibinfo{author}{\bibfnamefont{A.}~\bibnamefont{Zunger}},
  \bibinfo{journal}{Phys. Rev. B} \textbf{\bibinfo{volume}{23}},
  \bibinfo{pages}{5048} (\bibinfo{year}{1981}).

\bibitem[{\citenamefont{Perdew et~al.}(1996)\citenamefont{Perdew, Burke, and
  Ernzerhof}}]{pbe}
\bibinfo{author}{\bibfnamefont{J.~P.} \bibnamefont{Perdew}},
  \bibinfo{author}{\bibfnamefont{K.}~\bibnamefont{Burke}}, \bibnamefont{and}
  \bibinfo{author}{\bibfnamefont{M.}~\bibnamefont{Ernzerhof}},
  \bibinfo{journal}{Phys.\ Rev.\ Lett.} \textbf{\bibinfo{volume}{77}},
  \bibinfo{pages}{3865} (\bibinfo{year}{1996}).

\bibitem[{\citenamefont{Anisimov et~al.}(1991)\citenamefont{Anisimov, Zaanen,
  and Andersen}}]{ldaplusU}
\bibinfo{author}{\bibfnamefont{V.~I.} \bibnamefont{Anisimov}},
  \bibinfo{author}{\bibfnamefont{J.}~\bibnamefont{Zaanen}}, \bibnamefont{and}
  \bibinfo{author}{\bibfnamefont{O.~K.} \bibnamefont{Andersen}},
  \bibinfo{journal}{Phys. Rev. B} \textbf{\bibinfo{volume}{44}},
  \bibinfo{pages}{943} (\bibinfo{year}{1991}).

\bibitem[{\citenamefont{Dudarev et~al.}(1998)\citenamefont{Dudarev, Botton,
  Savrasov, Humphreys, and Sutton}}]{ldaplusU_simplified}
\bibinfo{author}{\bibfnamefont{S.~L.} \bibnamefont{Dudarev}},
  \bibinfo{author}{\bibfnamefont{G.~A.} \bibnamefont{Botton}},
  \bibinfo{author}{\bibfnamefont{S.~Y.} \bibnamefont{Savrasov}},
  \bibinfo{author}{\bibfnamefont{C.~J.} \bibnamefont{Humphreys}},
  \bibnamefont{and} \bibinfo{author}{\bibfnamefont{A.~P.}
  \bibnamefont{Sutton}}, \bibinfo{journal}{Phys. Rev. B}
  \textbf{\bibinfo{volume}{57}}, \bibinfo{pages}{1505} (\bibinfo{year}{1998}).

\bibitem[{\citenamefont{Youn et~al.}(2004)\citenamefont{Youn, Min, and
  Freeman}}]{mn_calco}
\bibinfo{author}{\bibfnamefont{S.~J.} \bibnamefont{Youn}},
  \bibinfo{author}{\bibfnamefont{B.~I.} \bibnamefont{Min}}, \bibnamefont{and}
  \bibinfo{author}{\bibfnamefont{A.~J.} \bibnamefont{Freeman}},
  \bibinfo{journal}{Phys. Stat. Sol. B} \textbf{\bibinfo{volume}{241}},
  \bibinfo{pages}{1411} (\bibinfo{year}{2004}).

\bibitem[{\citenamefont{Kunes et~al.}(2005)\citenamefont{Kunes, Ku, and
  Pickett}}]{eu_calco}
\bibinfo{author}{\bibfnamefont{J.}~\bibnamefont{Kunes}},
  \bibinfo{author}{\bibfnamefont{W.}~\bibnamefont{Ku}}, \bibnamefont{and}
  \bibinfo{author}{\bibfnamefont{W.}~\bibnamefont{Pickett}},
  \bibinfo{journal}{J. Phys. Soc. Jap.} \textbf{\bibinfo{volume}{74}},
  \bibinfo{pages}{1408} (\bibinfo{year}{2005}).

\bibitem[{\citenamefont{Giannozzi and et~al.}(2009)}]{QE}
\bibinfo{author}{\bibfnamefont{P.}~\bibnamefont{Giannozzi}} \bibnamefont{and}
  \bibinfo{author}{\bibnamefont{et~al.}}, \bibinfo{journal}{J. Phys.:Condens.
  Matter} \textbf{\bibinfo{volume}{21}}, \bibinfo{pages}{395502}
  (\bibinfo{year}{2009}).

\bibitem[{\citenamefont{Ramer and Rappe}(1999)}]{designed-nonlocal}
\bibinfo{author}{\bibfnamefont{N.~J.} \bibnamefont{Ramer}} \bibnamefont{and}
  \bibinfo{author}{\bibfnamefont{A.~M.} \bibnamefont{Rappe}},
  \bibinfo{journal}{Phys. Rev. B} \textbf{\bibinfo{volume}{59}},
  \bibinfo{pages}{12471} (\bibinfo{year}{1999}).

\bibitem[{opi()}]{opium}
\bibinfo{howpublished}{http://opium.sourceforge.net}.

\bibitem[{\citenamefont{Kresse and Hafner}(1993)}]{vasp}
\bibinfo{author}{\bibfnamefont{G.}~\bibnamefont{Kresse}} \bibnamefont{and}
  \bibinfo{author}{\bibfnamefont{J.}~\bibnamefont{Hafner}},
  \bibinfo{journal}{Phys. Rev. B} \textbf{\bibinfo{volume}{47}},
  \bibinfo{pages}{R558} (\bibinfo{year}{1993}).

\bibitem[{\citenamefont{Kresse and Furthmuller}(1996)}]{vasp2}
\bibinfo{author}{\bibfnamefont{G.}~\bibnamefont{Kresse}} \bibnamefont{and}
  \bibinfo{author}{\bibfnamefont{J.}~\bibnamefont{Furthmuller}},
  \bibinfo{journal}{Phys. Rev. B} \textbf{\bibinfo{volume}{54}},
  \bibinfo{pages}{11169} (\bibinfo{year}{1996}).

\bibitem[{\citenamefont{Bl\"ochl}(1994)}]{paw}
\bibinfo{author}{\bibfnamefont{P.~E.} \bibnamefont{Bl\"ochl}},
  \bibinfo{journal}{Phys. Rev. B} \textbf{\bibinfo{volume}{50}},
  \bibinfo{pages}{17953} (\bibinfo{year}{1994}).

\bibitem[{\citenamefont{Kresse and Joubert}(1999)}]{paw2}
\bibinfo{author}{\bibfnamefont{G.}~\bibnamefont{Kresse}} \bibnamefont{and}
  \bibinfo{author}{\bibfnamefont{D.}~\bibnamefont{Joubert}},
  \bibinfo{journal}{Phys. Rev. B} \textbf{\bibinfo{volume}{59}},
  \bibinfo{pages}{1758} (\bibinfo{year}{1999}).

\bibitem[{\citenamefont{Mostofi
  et~al.}(2008{\natexlab{a}})\citenamefont{Mostofi, Yates, Lee, Souza,
  Vanderbilt, and Marzari}}]{mlwf}
\bibinfo{author}{\bibfnamefont{A.~A.} \bibnamefont{Mostofi}},
  \bibinfo{author}{\bibfnamefont{J.~R.} \bibnamefont{Yates}},
  \bibinfo{author}{\bibfnamefont{Y.-S.} \bibnamefont{Lee}},
  \bibinfo{author}{\bibfnamefont{I.}~\bibnamefont{Souza}},
  \bibinfo{author}{\bibfnamefont{D.}~\bibnamefont{Vanderbilt}},
  \bibnamefont{and} \bibinfo{author}{\bibfnamefont{N.}~\bibnamefont{Marzari}},
  \bibinfo{journal}{Comput.\ Phys.\ Commun.} \textbf{\bibinfo{volume}{178}},
  \bibinfo{pages}{685} (\bibinfo{year}{2008}{\natexlab{a}}).

\bibitem[{\citenamefont{Mostofi
  et~al.}(2008{\natexlab{b}})\citenamefont{Mostofi, Yates, Lee, Souza,
  Vanderbilt, and Marzari}}]{wannier90}
\bibinfo{author}{\bibfnamefont{A.~A.} \bibnamefont{Mostofi}},
  \bibinfo{author}{\bibfnamefont{J.~R.} \bibnamefont{Yates}},
  \bibinfo{author}{\bibfnamefont{Y.-S.} \bibnamefont{Lee}},
  \bibinfo{author}{\bibfnamefont{I.}~\bibnamefont{Souza}},
  \bibinfo{author}{\bibfnamefont{D.}~\bibnamefont{Vanderbilt}},
  \bibnamefont{and} \bibinfo{author}{\bibfnamefont{N.}~\bibnamefont{Marzari}},
  \bibinfo{journal}{Comput. Phys. Commun.} \textbf{\bibinfo{volume}{178}},
  \bibinfo{pages}{685} (\bibinfo{year}{2008}{\natexlab{b}}).

\bibitem[{\citenamefont{Wang et~al.}(2006)\citenamefont{Wang, Yates, , Souza,
  and Vanderbilt}}]{qah_wan}
\bibinfo{author}{\bibfnamefont{X.}~\bibnamefont{Wang}},
  \bibinfo{author}{\bibfnamefont{J.~R.} \bibnamefont{Yates}}, ,
  \bibinfo{author}{\bibfnamefont{I.}~\bibnamefont{Souza}}, \bibnamefont{and}
  \bibinfo{author}{\bibfnamefont{D.}~\bibnamefont{Vanderbilt}},
  \bibinfo{journal}{Phys. Rev. B} \textbf{\bibinfo{volume}{74}},
  \bibinfo{pages}{195118} (\bibinfo{year}{2006}).

\bibitem[{\citenamefont{Wang et~al.}(2007)\citenamefont{Wang, Vanderbilt,
  Yates, and Souza}}]{qah_fermi}
\bibinfo{author}{\bibfnamefont{X.}~\bibnamefont{Wang}},
  \bibinfo{author}{\bibfnamefont{D.}~\bibnamefont{Vanderbilt}},
  \bibinfo{author}{\bibfnamefont{J.~R.} \bibnamefont{Yates}}, \bibnamefont{and}
  \bibinfo{author}{\bibfnamefont{I.}~\bibnamefont{Souza}},
  \bibinfo{journal}{Phys. Rev. B} \textbf{\bibinfo{volume}{76}},
  \bibinfo{pages}{195109} (\bibinfo{year}{2007}).

\bibitem[{\citenamefont{Souza}()}]{ivo_private}
\bibinfo{author}{\bibfnamefont{I.}~\bibnamefont{Souza}},
  \bibinfo{howpublished}{private communication}.

\end{thebibliography}

\end{document}